\documentclass[amsmath,amssymb,aps,pra,twocolumn,superscriptaddress]{revtex4-1}

\usepackage{graphicx}
\usepackage{dcolumn}
\usepackage{bm}
\usepackage{enumitem}
\usepackage{textcomp}
\usepackage{upgreek}

\begin{document}

\preprint{CEI_CO_AM}

\title{Coulomb explosion imaging of carbon monoxide dimers}

\author{A. M\'ery} \affiliation{CIMAP, CEA-CNRS-ENSICAEN-UNICAEN, Normandie Universit\'e,\\ BP5133, F-14050 Caen Cedex 04, France}
\author{V. Kumar} \affiliation{CIMAP, CEA-CNRS-ENSICAEN-UNICAEN, Normandie Universit\'e,\\ BP5133, F-14050 Caen Cedex 04, France}
\author{X. Fl\'echard} \affiliation{Normandie Univ, ENSICAEN, UNICAEN, CNRS/IN2P3, LPC Caen, 14000 Caen, France}
\author{B. Gervais} \affiliation{CIMAP, CEA-CNRS-ENSICAEN-UNICAEN, Normandie Universit\'e,\\ BP5133, F-14050 Caen Cedex 04, France}
\author{S. Guillous} \affiliation{CIMAP, CEA-CNRS-ENSICAEN-UNICAEN, Normandie Universit\'e,\\ BP5133, F-14050 Caen Cedex 04, France}
\author{M. Lalande} \affiliation{CIMAP, CEA-CNRS-ENSICAEN-UNICAEN, Normandie Universit\'e,\\ BP5133, F-14050 Caen Cedex 04, France}
\author{J. Rangama} \affiliation{CIMAP, CEA-CNRS-ENSICAEN-UNICAEN, Normandie Universit\'e,\\ BP5133, F-14050 Caen Cedex 04, France}
\author{W. Wolff} \affiliation{Instituto de F\'isica - Universidade Federal do Rio de Janeiro, Cidade Universit\'aria, Rio de Janeiro, Brazil}
\author{A. Cassimi} \affiliation{CIMAP, CEA-CNRS-ENSICAEN-UNICAEN, Normandie Universit\'e,\\ BP5133, F-14050 Caen Cedex 04, France}

\date{\today}

\begin{abstract}
We report on experimental results obtained from collisions of slow highly charged $\rm Ar^{9+}$ ions with a carbon monoxide dimer $\rm (CO)_2$ target. A COLTRIMS setup and a Coulomb explosion imaging approach are used to reconstruct the structure of the $\rm CO$ dimers. The three dimensional structure is deduced from the 2-body and 3-body dissociation channels from which both the intermolecular bond length and the relative orientation of the two molecules are determined. For the 3-body channels, the experimental data are interpreted with the help of a classical model in which the trajectories of the three emitted fragments are numerically integrated. We measured the equilibrium intermolecular distance to be $\rm R_e=4.2~\text{\AA}$. The orientation of both CO molecules with respect to the dimer axis is found to be quasi-isotropic due to the large vibrational temperature of the gas jet.
\end{abstract}



\maketitle

\section{INTRODUCTION}
\begin{figure}[htpb]
\includegraphics[scale=0.06]{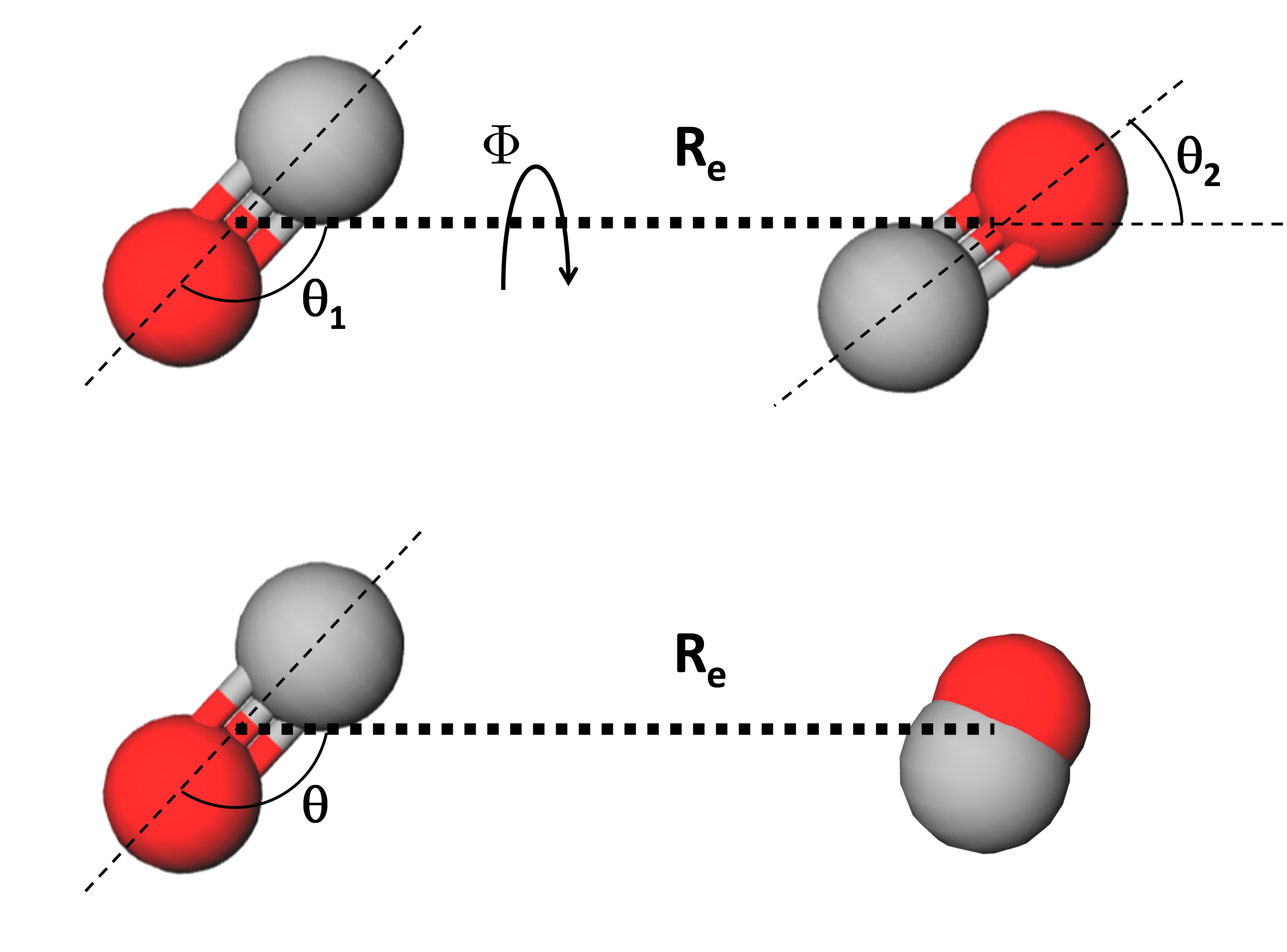}
\caption{\label{fig:Geom_coord}{Definition of the dimer geometry using internal coordinates. The CO molecular bond length is fixed to $\rm R_{CO}=1.13~\text{\AA}$. Upper panel: complete three-dimensional geometry ($\rm \uptheta_1$, $\rm \uptheta_2$, $\rm \Phi$, $\rm R_e$). Lower panel: Partial geometry accessible through the 2-body and 3-body dissociation of the dimer ($\rm \uptheta$, $\rm R_e$). The non-dissociating $\rm CO^+$ molecular ion is considered as a point-like particle located at its center of mass and the measurement is not sensitive to the $\rm \Phi$ coordinate (see details in the text).}}
\end{figure}

Carbon monoxide dimers $\rm (CO)_2$ have been studied since the 90$'$s using infrared and millimeter-wave spectroscopy [\onlinecite{BrokesJChemPhys1999},\onlinecite{Surin2003},\onlinecite{Rezaei2013}]. 
However, the interpretation of the measured rovibrational spectra has been puzzling for many years because of the possible contribution of two distinct isomers of the dimer with very little binding energy difference. The lowest in energy has been assigned to a planar C-bonded geometry in which the carbon atoms of each CO molecule are located in the inner region of the dimer. In contrast, the other isomer would correspond to a planar O-bonded configuration with two inner oxygen atoms. Moreover, these two isomers have been found to have different equilibrium intermolecular distance $\rm R_e$. The corresponding three-dimensional structures of the dimer can be described using four independent parameters: the intermolecular bond length $\rm R_e$, the relative orientation $\rm \uptheta_1$ and $\rm \uptheta_2$ of each molecule with respect to the dimer axis and the torsional angle $\rm \Phi$ (see Fig.~\ref{fig:Geom_coord}). Table~\ref{tab:table_geom_theo} shows a summarized comparison of the geometry of the C-bonded and O-bonded structures.

Conjointly, theoretical efforts were dedicated to the detailed investigation of the potential energy surface of the neutral CO dimer [\onlinecite{HanJMS1997},\onlinecite{Surin2007},\onlinecite{Dawes2013}]. Han et al. [\onlinecite{HanJMS1997}] reported about the possible existence of T-shaped structures (corresponding to $\rm \uptheta_1 \approx 0^{\circ}$ and $\rm \uptheta_2 \approx 90^{\circ}$) as well as slipped anti-parallel structures. The likely presence of planar anti-parallel geometries has been confirmed more recently through extensive efforts in the determination of the potential energy surface (PES) of CO dimers [\onlinecite{Surin2007},\onlinecite{Dawes2013}]. The PES shows a global minimum corresponding to the C-bonded geometry and a local minimum corresponding to the O-bonded geometry (see table ~\ref{tab:table_geom_theo}). These two stable structures differ both by their relative orientation $\rm \uptheta_1$ and $\rm \uptheta_2$ but also by their intermolecular distance $\rm R_e$ and thus may be identified thanks to two distinct experimental observables. The energies of these two minima have been calculated to be about -135 $\rm cm^{-1}$ and -120 $\rm cm^{-1}$ respectively. The two potential wells are separated by low barriers following a dis-rotatory path along the $\rm X = \frac{\uptheta_1+\uptheta_2}{2}$ coordinate while passing through some intermediate parallel conformations. The low barrier (about 18 $\rm cm^{-1}\approx 2 ~meV$) easily enables vibrational excitation along this dis-rotatory coordinate [\onlinecite{Dawes2013}] even for relatively low excitation energies of the dimer. 

The Coulomb explosion imaging (CEI) technique is known to be a powerful tool to investigate the fragmentation dynamics of covalent molecules [\onlinecite{LegarePRA2005},\onlinecite{MatsudaJCP2007},\onlinecite{NeumannPRL2010},\onlinecite{WuJCP2015}]. The method has also recently allowed the experimental determination of the structure of van der Waals complexes such as rare gas trimers [\onlinecite{UlrichJPC2011}] but also mixed $\rm ArN_2$ [\onlinecite{WuJCP2014}] or $\rm ArCO$ [\onlinecite{GongPRA2013}] dimers. The CEI technique is particularly well suited for van der Waals clusters where large intermolecular distances lead to dissociation following a pure Coulomb potential energy curve [\onlinecite{WuPRL2013}]. This experimental method also requires a rapid ionization of the target molecules in such a way that the nuclear motion can be considered as frozen during the ionization process.

The CEI technique gives access to several experimental observables, such as the kinetic energy released (KER) in the fragmentation and the relative angles of emission of the fragments, which provide unique information on the 3D molecular structure. It is worth noting that the direct measurement of the complete structure of a carbon monoxide dimer using the CEI technique would require the Coulomb explosion of the two constitutive CO molecules and the coincident detection of the four emitted atomic ions. Although, the cross sections for creation of 4-fold ionized dimers leading to such 4-body fragmentation channels are very small in the present collision system. Moreover, the data analysis procedure is complex because of the strong ambiguity for pairing two sets of carbon and oxygen atomic ions coming from two identical molecules. Thus, only the 2-body and 3-body dissociation channels will be presented in the forthcoming discussion. As described later, the 2-body channels (where only intermolecular fragmentation occurs) give a direct measurement of the intermolecular distance while the molecular orientation can be deduced from the 3-body channels. In the following we will refer to the molecular orientation using a single angle $\rm \uptheta$ because the fragmentation in three bodies does not enable measuring both $\rm \uptheta_1$ and $\rm \uptheta_2$ and neither the torsional angle $\rm \Phi$ (Fig.~\ref{fig:Geom_coord}, lower panel). Note that the $\rm \uptheta$ angle is defined with respect to the van der Waals bond and the oxygen atom. As a result, the O-bonded structure corresponds to $\rm \uptheta=45^{\circ}$ while the C-bonded structure is associated to $\rm \uptheta=135^{\circ}$.

\begin{table}
\begin{ruledtabular}
\begin{tabular}{ccccc}
\textrm{Structure} 				& \multicolumn{2}{c}{C-bonded} 							& \multicolumn{2}{c}{O-bonded} \\
\textrm{} 								& \textrm{expt.} & \textrm{calc.} 						& \textrm{expt.} & \textrm{calc.} \\
\textrm{} 								& [\onlinecite{BrokesJChemPhys1999}] & [\onlinecite{Surin2007},\onlinecite{Dawes2013}] & [\onlinecite{BrokesJChemPhys1999}] & [\onlinecite{Surin2007},\onlinecite{Dawes2013}] \\
\colrule
$\rm{\uptheta (^{\circ})}$ 	& --- 		& 134.2/134.79 										& --- 		& 61/65.18 \\
$\rm{R_e (\text{\AA})}$ 	&	4.4 		&	4.33/4.331 											&	4.0 	&	3.67/3.620 \\
\end{tabular}
\end{ruledtabular}
\caption{\label{tab:table_geom_theo} 
Structural parameters for the planar C-bonded and O-bonded geometries of the neutral CO dimer. The relative orientation angle $\rm \uptheta$ with respect to the dimer axis and the intermolecular distance are defined on Fig.~\ref{fig:Geom_coord}.
}
\end{table}

In the present study, multiply charged dimers are produced following multiple electron capture by slow highly charged ions on neutral dimer targets. The velocity of the $\rm Ar^{9+}$ projectiles is about $\rm 8~\text{\AA} /fs$ (0.37 atomic units) and the geometrical size of the dimer target is of the order of $\rm 4.5~\text{\AA}$. For such projectile ions, electron capture is by far the dominant process and occurs in the femtosecond timescale. The subsequent concerted fragmentation of the ionized target also occurs on the femtosecond timescale and ensures the validity of the axial recoil approximation on which relies our Coulomb explosion approach. We focus here on collisions where at least one electron is captured on each molecular site of the CO dimer thus leading to the cleavage of the intermolecular van der Waals bond. Covalent bond breakup also occurs when two or more electrons are captured on one of the CO molecules. The relevant fragmentation channels can be classified into 2-body or 3-body channels depending on either bound or dissociative states of the multi-ionized CO molecule are formed during the collision process. The following 2-body channels have been experimentally identified and are referred to as (m,n) where m and n are the charge states of the emitted CO molecular ions:
\begin{center}
(1,1) :  $\rm (CO)_2^{2+} \rightarrow CO^+ + CO^+$ \\
\end{center}

\begin{center}
(2,1) :  $\rm (CO)_2^{3+} \rightarrow CO^{2+} + CO^+$ \\
\end{center}

The simultaneous break-up of both the van der Waals bond and of one covalent bond leads to the following 3-body channels:
\begin{center}
	(1,[1,1]) :  $\rm (CO)_2^{3+} \rightarrow CO^+ + C^+ + O^+$ \\
	\end{center}

\begin{center}
	(1,[2,1]) :  $\rm (CO)_2^{4+} \rightarrow CO^+ + C^{2+} + O^+$ \\
	\end{center}

\begin{center}
	(1,[1,2]) :  $\rm (CO)_2^{4+} \rightarrow CO^+ + C^+ + O^{2+}$ \\
	\end{center}

\begin{center}
	(1,[2,2]) :  $\rm (CO)_2^{5+} \rightarrow CO^+ + C^{2+} + O^{2+}$ \\
\end{center}

These channels are referred to as (m,[p,q]) where m=1 is the charge state of the non-dissociating $\rm CO^+$ cation and where p and q are respectively the charge states of the carbon and oxygen atomic ions. The simultaneous measurement of several 2-body and 3-body fragmentation channels allows an unambiguous determination of the $\rm R_e$ and $\rm \uptheta$ coordinates of the CO dimer from the present data set.

\section{EXPERIMENT}
The experiment uses the COLd Target Recoil Ion Momentum Spectroscopy technique to perform the coincident detection of the positively charged fragments emitted from collisions of CO dimers with $\rm Ar^{9+}$ projectile ions. Both target and projectile beams cross at right angle at the center of the spectrometer. A homogeneous perpendicular electric field of about 40~V/cm is oriented along the vertical X axis and allows the collection of the cations resulting from fragmentation on a micro-channel plate detector coupled to delay line anodes. The $\rm 4\pi$ collection efficiency is restricted by the 80~mm diameter of the recoil ion detector located 100~mm from the collision point resulting in a limit of 16~eV (resp. 32~eV) for the detection of singly (resp. doubly) charged ions. Before reaching the detector the ions are further post-accelerated to about 3~keV per charge to ensure an optimal detection efficiency of the MCP. Finally, the three dimensional coordinates of the momentum vector of each fragment are computed from their time-of-flight and position on the detector. 
The highly charged $\rm Ar^{9+}$ projectile ions are extracted from an electron cyclotron resonant ion source at a kinetic energy of 135 keV on the ARIBE/GANIL beam line. The ion beam then passes through a $\rm 600~\mu m$ collimation aperture before colliding with the gas jet. At the exit of the spectrometer, the final charge state of the projectile ions is analyzed thanks to a parallel plates deflector before reaching a dedicated position sensitive detector. Argon projectiles with charge state ranging from 5+ to 8+ give the start for the time-of-flight measurement while the intense primary $\rm Ar^{9+}$ beam is deflected out of the detector. No selection is imposed on the final charge state of the projectile ion because it shows no influence on the fragmentation dynamics of the ionized target.

The neutral CO dimer targets are produced from a supersonic gas jet with a stagnation pressure of 15 bars, expending through a $\rm 30~\mu m$ nozzle. A skimmer of 0.5 mm diameter is located 6 mm away from the nozzle and a collimator of 0.5 mm diameter is placed before the collision chamber. The resulting diameter of the gas jet at the collision point is about 1.5~mm. The translational and rotational temperatures of the gas jet are estimated to be below 10~K. However, the vibrational degrees of freedom are only partially cooled down in the supersonic expansion and the vibrational temperature of the jet is expected to be close to the temperature of the nozzle which is here 300~K [\onlinecite{McClellandJPC1979},\onlinecite{WallJPB2016}]. The proportion of CO dimers within the gas jet is estimated to be about 1~\% of the dominant monomer targets and the yield of larger cluster $\rm (CO)_{n\geq3}$ targets is at least one order of magnitude lower than the dimers. As a consequence, an accurate data filtering method is mandatory to extract the relatively small portion of events originating from the relevant dimer break-ups. For this purpose, the coincident measurement of all emitted cationic fragments and the restriction imposed by the momentum conservation law enable to efficiently select the desired fragmentation channels.

\begin{figure}[htpb]
\includegraphics[scale=0.11]{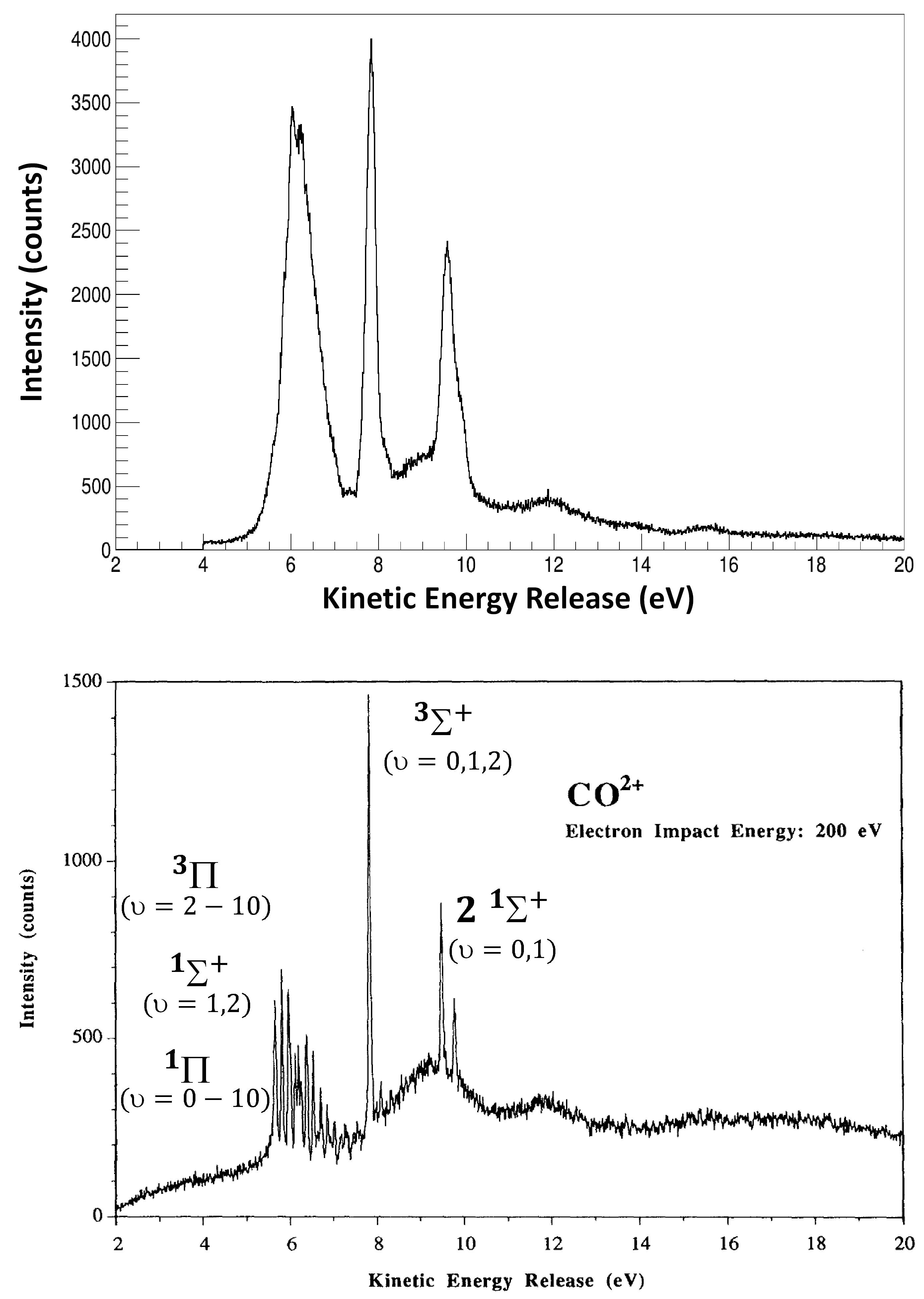}
\caption{\label{fig:KER_dication}{KER spectra of the monomer dication dissociation $\rm CO^{2+} \rightarrow C^+ + O^+$. Upper panel: KER spectrum from the present measurement. Lower panel: KER spectrum from electron impact ionization measurement [\onlinecite{Lundqvist1995}] and identification of the corresponding electronic states of the $\rm CO^{2+}$ dication [\onlinecite{Lundqvist1995}]. The relative deviation from the reference data is below 1 \% over the 7 to 10 eV energy range.}}
\end{figure}

In the framework of the Coulomb imaging technique, the intermolecular bond length $\rm R_e$ is derived from the KER measurement of the 2-body dissociation channels. Therefore, special attention has been devoted to the absolute momentum calibration of the spectrometer. This calibration requires both time-of-flight and position calibrations along the two transverse directions Y and Z of the spectrometer. The time-of-flight (TOF) calibration is performed by using the centroids of several well identified peaks in the TOF spectrum resulting from the detection of different atomic or molecular ions and allows a precise determination of the spectrometer characteristics (flight length and electric field strength). Additionally, the measured TOF are further corrected to first order from the short flight time of the recoil ions in the post acceleration region. The position calibration along the Y and Z directions of the delay line anodes was adjusted to get the optimal KER resolution, which is only achieved when the three components of each fragment momentum are properly reconstructed. We used the dominant $\rm CO^{2+} \rightarrow C^+ + O^+$ fragmentation of the monomer dication and the reference KER spectrum obtained with a Doppler free instrument [\onlinecite{Lundqvist1995}] for tuning the absolute energy calibration of our spectrometer (Fig.~\ref{fig:KER_dication}). Even if our energy resolution does not allow to separate the vibrational levels, the KER peak associated to the $\rm ^3\Sigma^+$ state (KER=7.833 eV [\onlinecite{Lundqvist1995}]) can be used to test the absolute calibration of our apparatus as it contains mainly one vibrational level. Our experimental energy resolution has also been estimated from the width of this KER peak and is found to be $\rm \Delta E$=0.28~eV (FWHM) for a KER of 7.833~eV. Moreover, the contributions of two vibrational levels associated to the $\rm 2 ^1\Sigma^+$ state (Fig.~\ref{fig:KER_dication}) have also been deconvoluted and fitted from the high energy peak around 9.5~eV. The results show that our KER calibration is consistent with the reference data from [\onlinecite{Lundqvist1995}] with a precision better than 1 \% over the 7~eV to 10~eV range. The resulting uncertainty on the intermolecular distance $\rm R_e$ is thus way below 10~\% and would allow separating the two predicted C-bonded and O-bonded structures of the CO dimer.

\section{EXPERIMENTAL RESULTS}

\subsection{ 2-body dissociation channels}
\begin{figure}[htpb]
\includegraphics[scale=0.11]{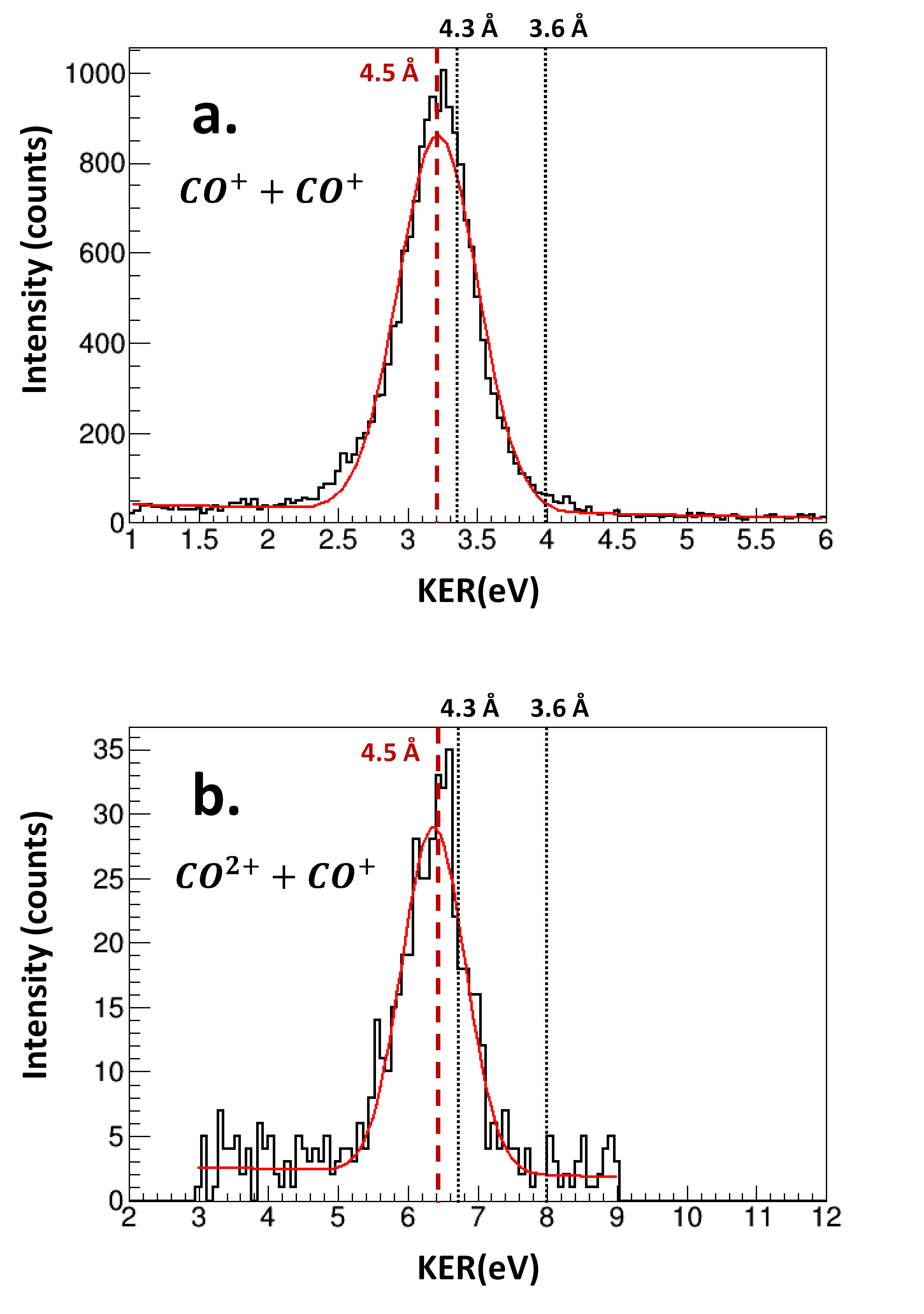}
\caption{\label{fig:KER_2bodie}{KER spectra of the 2-body a. (1,1) and b. (2,1) dissociation channels. For both spectra, the background contribution is due to random coincidences and is subtracted using a linear function and the red line is a Gaussian fit of the experimental data. The expected position of the KER peaks corresponding to theoretical predictions $\rm R_e=4.3~\text{\AA}$ (C-bonded) and $\rm R_e=3.6~\text{\AA}$ (O-bonded) are represented by dashed black lines.}}
\end{figure}

We first consider the Coulomb explosion of the dimer resulting from a single electron capture on each molecule $\rm (CO)_2^{2+} \rightarrow CO^+ + CO^+$. Fig.~\ref{fig:KER_2bodie}.a.) shows the associated KER spectrum with a single peak centered at $\rm KER^{(1,1)}=3.21 \pm 0.04~eV$. This value agrees very well with the former experimental measurement by Ding et al. [\onlinecite{DingPRL2017}] in which the similar dissociation channel of the $\rm CO$ dimers was triggered by multiple ionization with ultra-short intense laser pulses. The presence of a single peak in the KER spectrum indicates no sizable contribution of indirect relaxation processes such as ICD (Intermolecular Coulomb Decay) or RCT (Radiative Charge Transfer). The Coulomb explosion approximation can thus be used to infer the intermolecular distance from the KER spectrum. In this approximation, the equilibrium distance between the centers of mass of the two CO molecules is found to be $\rm R_e^{(1,1)}=4.49~\text{\AA}$.

However the $\rm CO^+$ ion has a quite large electrostatic dipole moment ($\rm \mu \approx 2.75 D$) and a significant part of the initial potential energy can thus be released as rotational energy. This rotational energy is not accounted for in the KER spectrum deduced from the $\rm CO^+$ fragment momenta and leads to an overestimation of the $\rm R_e$ value deduced from the pure Coulomb model. The correction of the KER measurement due to the undetected rotational energy will be detailed in section V.B and leads the actual intermolecular distance to be $\rm R_e^{(1,1)}=4.2~\text{\AA}$.

A small contribution of the $\rm (CO)_2^{3+} \rightarrow CO^{2+} + CO^+$ dissociation channel has also been identified (Fig.~\ref{fig:KER_2bodie}.b). Its low relative intensity results mainly from the low proportion of metastable states (with a lifetime longer than $\rm 1 \mu s$) of the $\rm CO^{2+}$ dication in comparison with the dissociative states leading to the competing 3-body channel $\rm (CO)_2^{3+} \rightarrow CO^+ + C^+ + O^+$. The KER peak is centered at $\rm KER^{(2,1)}=6.37\pm0.07 eV$ which is two times the KER measured in the $\rm CO^+ + CO^+$ channel as expected within the approximation of the Coulomb explosion. A correction from the rotational energy (see section V.B) leads to an intermolecular distance of $\rm R_e \cong 4.3~\text{\AA}$ for the $\rm CO^{2+} + CO^+$ channel.

The two intermolecular distances inferred from different dissociation channels ($\rm R_e \cong 4.2~\text{\AA}$) seems more likely to correspond to the C-bonded structure predicted by theoretical calculations (see table ~\ref{tab:table_geom_theo}). As already discussed in a previous section, this geometry has been calculated to be the global minimum of the PES of the neutral CO dimer and is associated to $\rm \uptheta_a=\uptheta_b=135^{\circ}$. The measured KER distributions of Fig.~\ref{fig:KER_2bodie} do not show any visible peak corresponding to the intermolecular distance of $3.6~\text{\AA}$ expected for the isomeric O-bonded structure. Contrary to previous IR measurements where several subbands were interpreted to arise from two different dimer bond lengths ($4.0$ and $4.4~\text{\AA}$), we observe here only one intermolecular distance. Our value is close to the calculated distance of the true ground state of the dimer with a C-bonded orientation. Further interpretation of this bond length measurement will be given in the next section especially in regards of the supersonic jet temperature.

\subsection{ 3-body dissociation channels}
In order to look for a more direct evidence for such a geometrical structure, the 3-body channels will now be investigated. We focus here on several particular 3-body dissociation channels (m=1,[p,q]) of the dimer with the aim to determine the $\rm \uptheta$ coordinate, i.e. the relative orientation of the CO molecules in the dimer. These channels result from a single electron capture on one molecule and a double, triple or quadruple electron capture on the other partner of the dimer. The use of the CEI technique requires that the 3-body break-up occurs simultaneously or at least on a smaller time scale than the relative motion of each fragment ion. Such a specific case will be later referred to as concerted fragmentation. Conversely, the existence of dissociative states of the multi-ionized CO molecule with lifetimes much longer than 10 ps will lead to a so-called sequential fragmentation. Such long-lived states do not exist for multiply charged $\rm (CO)^{n\geq3+}$ cations [\onlinecite{KumarJP2010}] and sequential fragmentation is thus expected to be observed only in the (1,[1,1]) channel.

\subsubsection{ The (1,[1,1]) channel: concerted and sequential fragmentation}
Since its first observation within a mass spectrometer in 1932 [\onlinecite{Friedlander1932}], the lifetime of metastable states of $\rm CO^{2+}$ has been widely studied both experimentally and theoretically [\onlinecite{Andersen1993}-\onlinecite{MrugalaJCP2008}]. When such metastable states are populated, the late fragmentation of the covalently bond molecule washes most of the information on the initial geometry. In order to identify these events and exclude them from the analysis, we will first study the Dalitz representation of the 3-body fragmentation.

\begin{figure}[htpb]
\includegraphics[scale=0.07]{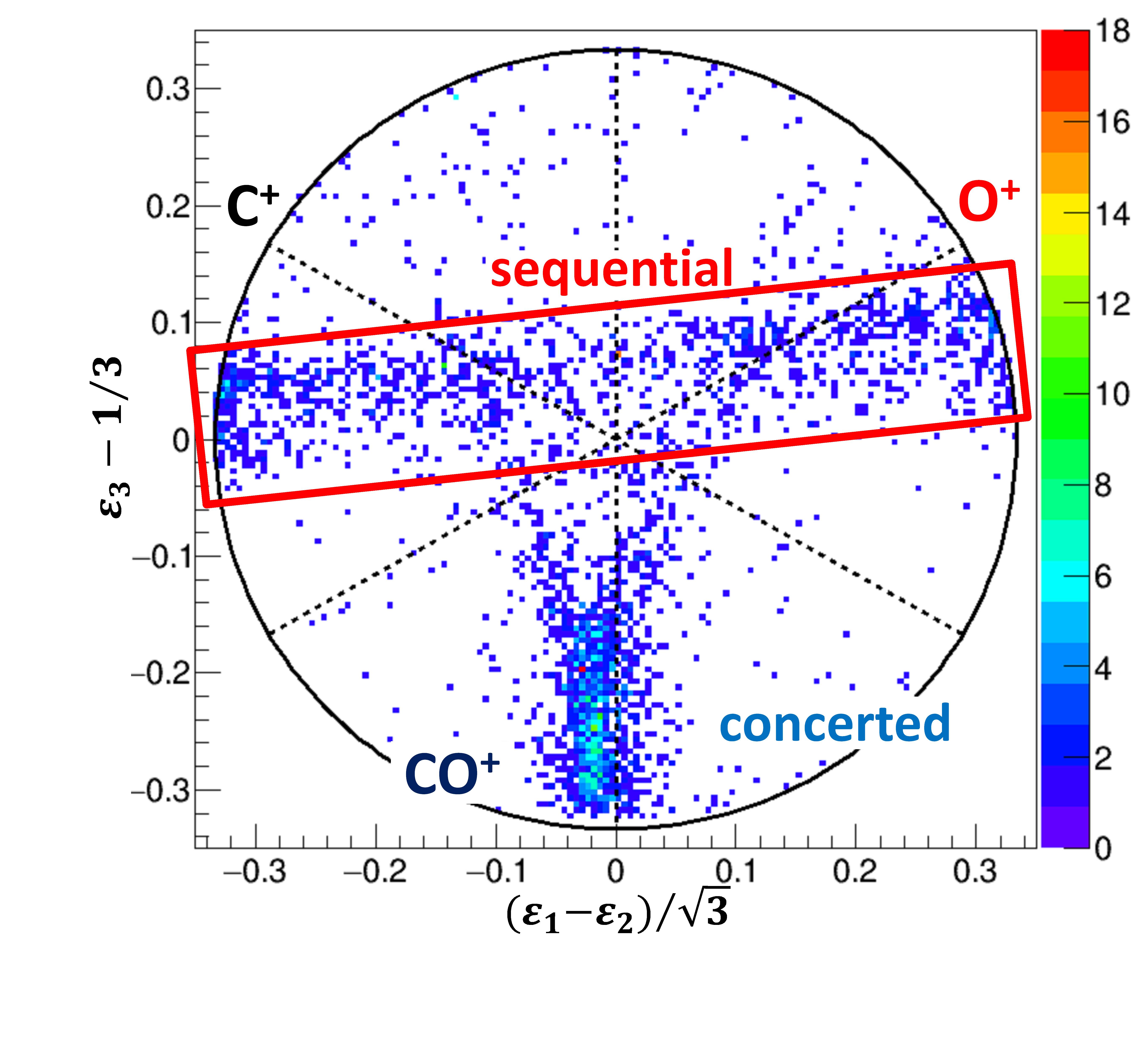}
\caption{\label{fig:Dalitz}{Dalitz plot of the (1,[1,1]) dissociation channel and identification of the concerted and sequential fragmentation processes.}}
\end{figure}

Fig.~\ref{fig:Dalitz} shows the Dalitz diagram of the 3-body dissociation of the dimer. This two-dimensional graph represents the reduced squared momenta of each fragment $\rm \varepsilon_i=\frac{p_i^2}{\sum _{i=1}^{3} p_i^2}$ (where the subscript \textit{i} stands respectively for the $\rm C^+$, $\rm O^+$ and $\rm CO^+$ recoiling ions) along the three axis of the diagram. The corresponding Cartesian coordinates are calculated as : $\rm x_D=\frac{\varepsilon_1 - \varepsilon_2}{\sqrt{3}}$ and $\rm y_D=\varepsilon_3 - \frac{1}{3}$ where \textit{i} is the number of the fragment sorted by ascending order of mass-to-charge ratio.
Two main contributions are clearly visible and are assigned respectively to a concerted or a sequential dissociation depending on the lifetime of the transient $\rm CO^{2+}$ dication [\onlinecite{DingPRL2017},\onlinecite{SongPRA2019}]. 
The quasi-horizontal line in the upper part of the graph is associated to the sequential fragmentation. In such a two-step process, the $\rm CO^{2+}$ dication breaks up only after the two recoiling molecular ions have reached the asymptotic limit for which their relative Coulomb repulsion vanishes. As shown in the next section (Fig.~\ref{fig:Simu_dynamics}.b), such a limit is reached after about 10 ps and only metastable states of $\rm CO^{2+}$ with lifetimes longer than 10 ps contribute to the sequential dissociation process. The recoiling $\rm CO^{2+}$ ion has thus enough time to rotate before fragmenting and completely loses the information on its initial orientation within the dimer. 
Contrarily, in a concerted fragmentation the $\rm CO^{2+}$ dication dissociates within a timescale shorter than 10 fs (see Fig.~\ref{fig:Simu_dynamics}.b). The corresponding events are located close to the bottom of the vertical axis. Here, the $\rm CO^+$ acquires a smaller kinetic energy because of the fast repulsion of the two $\rm C^+$ and $\rm O^+$ atomic ions following the dissociation of the $\rm CO^{2+}$ dication. The exact distribution of these events is directly correlated to the initial structure of the dimer but also to the $\rm CO^{2+}$ fragmentation dynamics. Only these concerted fragmentation events contain the desired information about the initial three-dimensional structure of the dimer. This lower part of the Dalitz distribution will later be compared to classical mechanics calculations with the aim to infer the relative molecular orientation in the dimer. Note that a "V shape" contribution is also visible on the Dalitz plot but with a very small intensity (expending from $\rm y_D\cong-0.2$ up to the upper part of the plot). These events are also assigned to a concerted fragmentation and will be discussed in the last section.

\subsubsection{The multiply charged (1,[p,q]) channels}
If one electron is captured on the first molecule and more than two electrons are captured from the other molecule, the dimer undergoes a concerted 3-body explosion as there are no long-lived states of the $\rm CO^{n+} (n\geq3)$ cations. In the present experiment, events corresponding to n=3 and n=4 have been recorded. The cross sections for multiple ionization decrease with the number of captured electrons and result in low statistics for these highly charged channels. 
For n=3, two distinct channels compete depending on the charge sharing between the emitted carbon and oxygen ions: (1,[1,2]) and (1,[2,1]). We found that the relative yield of the (1,[1,2]) is approximately five times lower than the dominant (1,[2,1]) channel. This ratio is compatible with previous measurements for monomer targets and results from the lower ionization potentials of the carbon ions [\onlinecite{BenItzhakPRA1993}].
For n=4, the symmetric (1,[2,2]) channel is found to be by far the dominant channel and no significant contribution of the asymmetric channels (1,[1,3]) and (1,[3,1]) has been observed. The preference for symmetric charge sharing in the fragmentation of the $\rm CO^{4+}$ ion has also been reported for monomers targets [\onlinecite{BenItzhakPRA1993}] and agrees with our measurement.

\begin{figure}[htpb]
\includegraphics[scale=0.18]{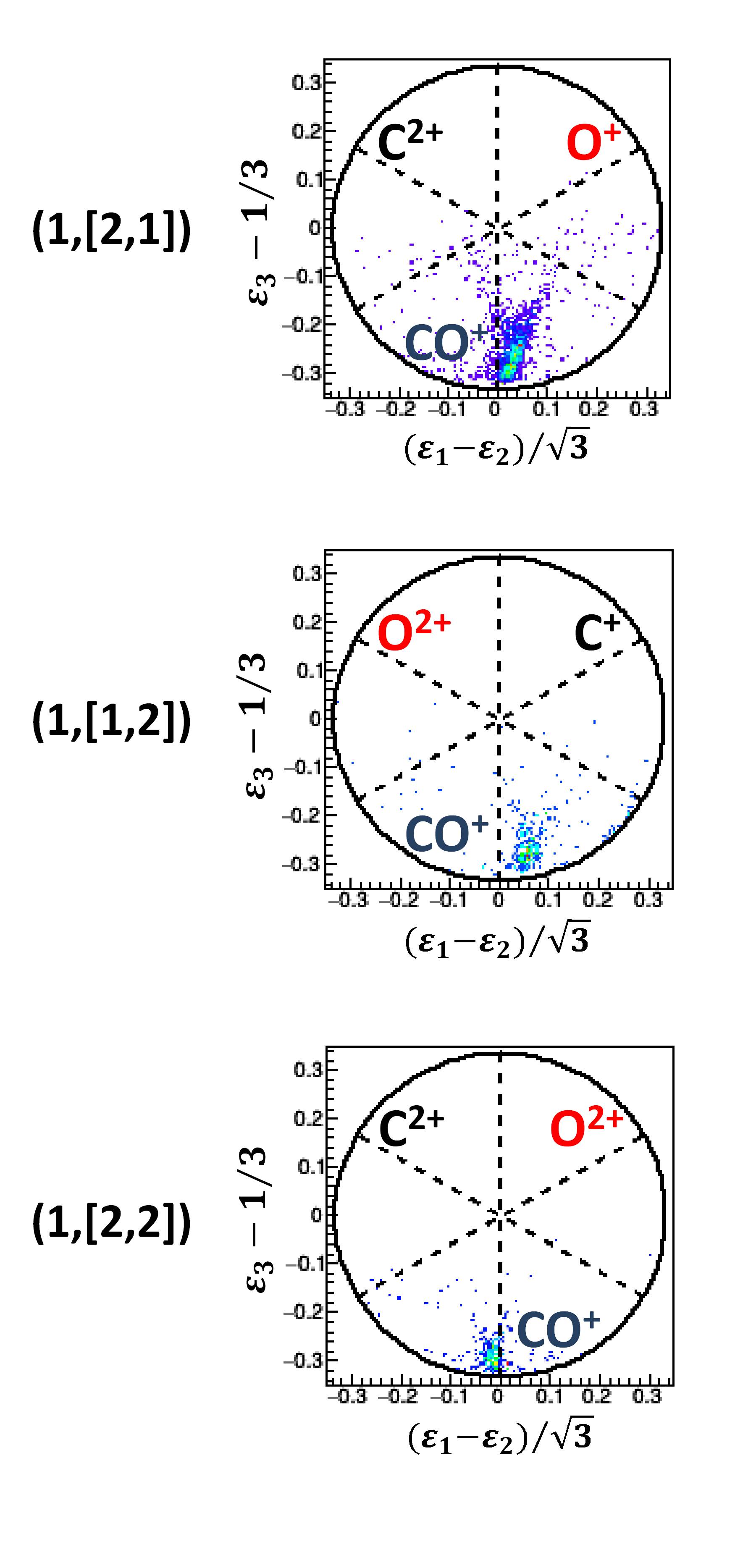}
\caption{\label{fig:Dalitz_multicharged}{Dalitz diagrams for the (1,[2,1]), (1,[1,2]) and (1,[2,2]) channels. The Cartesian coordinates are: $\rm x_D=\frac{\varepsilon_1 - \varepsilon_2}{\sqrt{3}}$ and $\rm y_D=\varepsilon_3 - \frac{1}{3}$ where $\rm \varepsilon_i (i=1, 2, 3)$ are the reduced squared momenta of the fragments, sorted by ascending order of mass-to-charge ratio. Using this convention, the momentum of the $\rm CO^+$ ion is plotted along the vertical axis but the carbon and oxygen ions are plotted the right or left diagonals depending on their respective charge states.}}
\end{figure}

The Dalitz plots for the (1,[1,2]), (1,[2,1]) and (1,[2,2]) channels are shown in Fig.~\ref{fig:Dalitz_multicharged}. For these channels, no sequential fragmentation is observed but a weak "V shape" contribution can be identified in the (1,[2,1]) channel. The main density distribution stands on the right or left side of the graph depending on the mass-over-charge ratio of each ion. A more detailed description of these graphs will be given when comparing to the simulated data in the next section.

\section{COMPUTATIONAL METHOD}
The momentum sharing in the 3-body break-up of the dimer has been investigated using Monte Carlo simulations based on a classical model where each charged fragment is considered as a point-like particle located at its center-of-mass. The complex fragmentation dynamics of the molecular $\rm CO^{n+} (n\geq 2)$ has been simply accounted for using the experimental KER spectra. For each simulated event, the classical trajectories of the three emitted ions are then numerically integrated to access the final momentum of each fragment. Despite its simplicity, the resulting model gives a good quantitative description of the dependence of the experimental observables on the geometrical structure of the dimer. 

\subsubsection{Description of the model}
The model assumes a concerted 3-body fragmentation in which both the van der Waals and the covalent bonds break at the same time. The three ions are considered as three point-like particles with zero initial velocity and are initially positioned accounting for the intermolecular distance $\rm R_e=4.5~\text{\AA}$, the molecular bond length $\rm R_{CO}=1.13~\text{\AA}$ of the neutral CO and the orientation $\rm \uptheta$ of the dissociating $\rm CO^{n+}$ with respect to the dimer axis (see Fig.~\ref{fig:Geom_coord}). Note that the qualitative description of the simulated Dalitz graphs is only weakly sensitive to the bond length and we used here the value $\rm R_e=4.5~\text{\AA}$ deduced from the pure 2-body Coulomb model (see section III.A). Then, the equations of motion are integrated using a standard Runge-Kutta method and assuming a pure Coulomb repulsion between both atomic $\rm C^{p+}$ and $\rm O^{q+}$ ions and the non-dissociating $\rm CO^+$ molecular ion. Contrarily, the mutual interaction between the $\rm C^{p+}$ and $\rm O^{q+}$ ions is known to be not purely Coulombic as it can be obviously demonstrated by the large KER distributions for these channels. Accounting for the KER distributions by the mean of \textit{ab-initio} calculations of the potential energy curves would be a tedious task. We thus use a scaled Coulomb energy curve directly inferred from the experimental data. For each simulated event, the total KER is first sampled randomly according to the experimental KER distribution. The 2-body KER for $\rm C^{p+}$ and $\rm O^{q+}$ ions is then obtained by subtracting the potential energy due to Coulomb interaction with the remaining $\rm CO^+$ molecular ion at a distance $\rm R_e$. The 2-body repulsive potential between the $\rm C^{p+}$ and $\rm O^{q+}$ ions is then approximated by : $\rm V_{CO}(r)=\frac{p \times q}{4 \pi \epsilon_0} \frac{A_{CO}}{r}$ where $\rm r$ is the distance between the $\rm C^{p+}$ and the $\rm O^{q+}$ ions and where the unit-less parameter $\rm A_{CO}$ is set to obtain the desired KER value. Such a potential curve gives rise to a 2-body KER of $\rm \frac{p \times q}{4 \pi \epsilon_0} \frac{A_{CO}}{R_{CO}} = p \times q \times A_{CO} \times 12.74 eV$ for the monomer cation as the bond length is fixed at $\rm R_{CO}=1.13~\text{\AA}$. The parameter is chosen either $\rm A_{CO}>1$ or $\rm A_{CO}<1$ in order to reproduce the experimental KER spectrum and typical values range from $\rm A_{CO}=0.4$ to $\rm A_{CO}=4$. The 3-dimensional coupled differential equations of motion are then deduced from the forces acting between each partner: 
	\begin{center}
	$\rm \left\lVert\overrightarrow{F_{12}}\right\rVert=\frac{p}{4 \pi \epsilon_0} \frac{1}{r_{12}} $\\
	\end{center}
	\begin{center}
	$\rm \left\lVert\overrightarrow{F_{13}}\right\rVert=\frac{q}{4 \pi \epsilon_0} \frac{1}{r_{13}} $\\
	\end{center}
	\begin{center}
	$\rm \left\lVert\overrightarrow{F_{23}}\right\rVert=\frac{p\cdot q}{4 \pi \epsilon_0} \frac{A_{CO}}{r_{23}} $\\
	\end{center}

where the subscript 1, 2 and 3 stand for the $\rm CO^+$, $\rm C^{p+}$ and $\rm O^{q+}$ ions respectively.

\begin{figure}[htpb]
\includegraphics[scale=0.08]{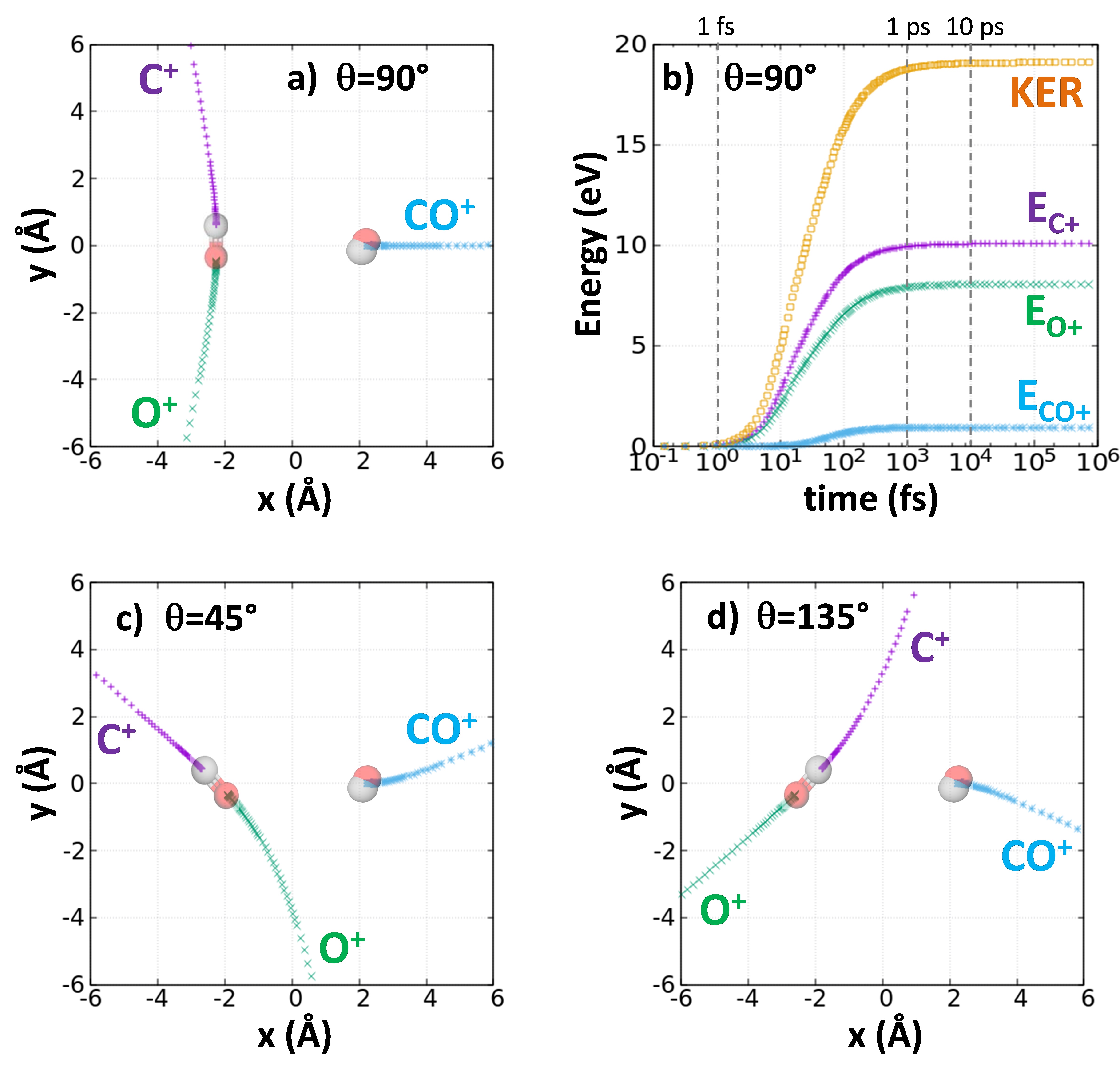}
\caption{\label{fig:Simu_dynamics}{a), c) and d) : calculated ion trajectories of the 3-body dissociation of the (1,[1,1]) channel for $\rm \uptheta =90^{\circ},45^{\circ},135^{\circ}$. b) : Evolution of the kinetic energy and total KER as a function of time for $\rm \uptheta =90^{\circ}$. The asymptotic limit is reached after about 10 ps. }}
\end{figure}

Finally, the numerical integration interval is set to 1 ns which is a sufficiently long time to ensure that all ions have reached their asymptotic momentum vector limit. Fig.~\ref{fig:Simu_dynamics} shows three examples of such a 3-body dissociation for the (1,[1,1]) channel, with $\rm A_{CO}=1$ but using different initial geometries $\rm \uptheta =90^{\circ},45^{\circ},135^{\circ}$. For all cases, all ions reach their kinetic energy limit after few picoseconds (see Fig.~\ref{fig:Simu_dynamics}.b).

\begin{figure}[htpb]
\includegraphics[scale=0.075]{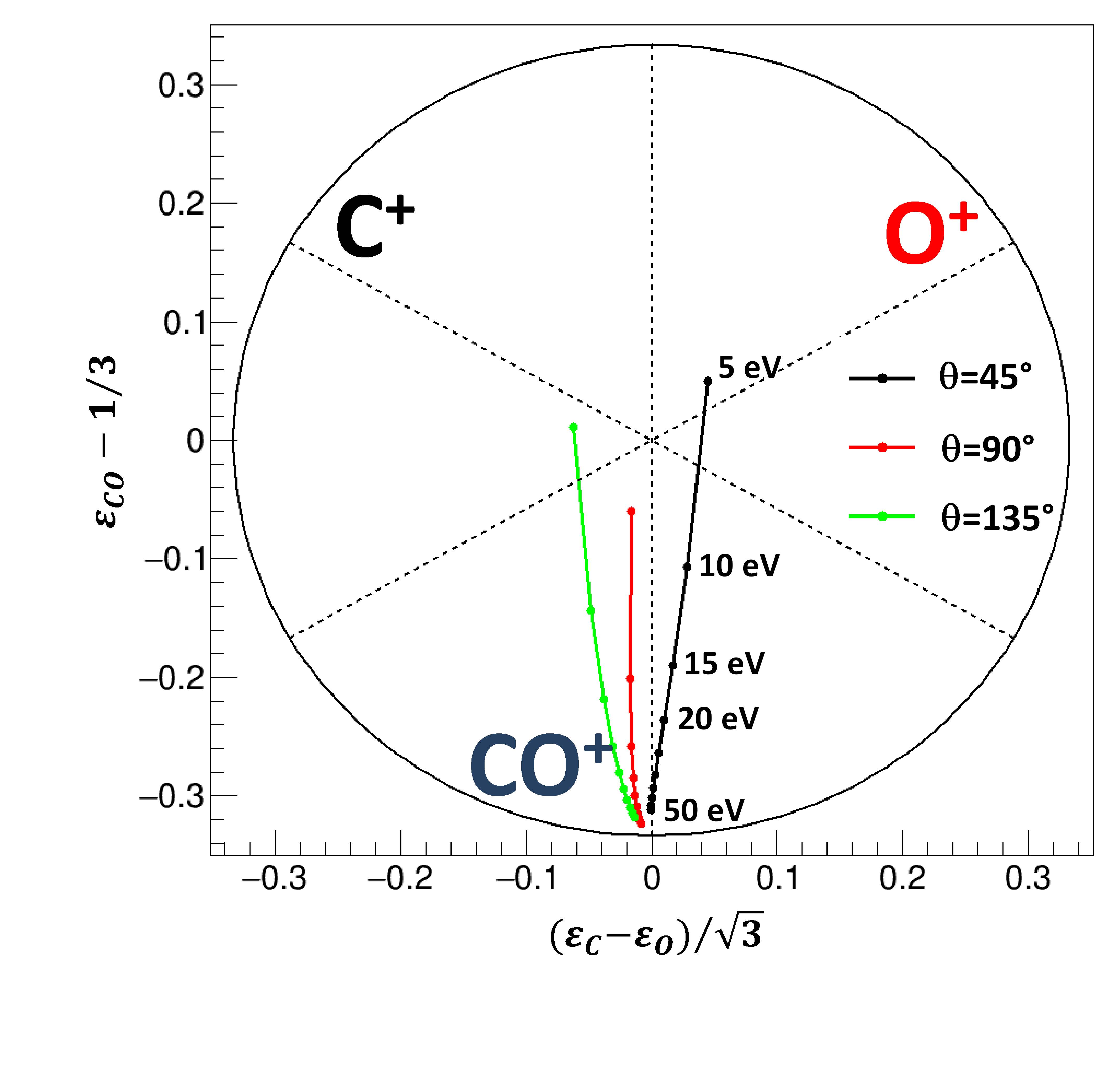}
\caption{\label{fig:Simu_Dalitz}{Dalitz representation of the reduced squared momenta in the concerted 3-body dissociation of the (1,[1,1]) channel for $\rm \uptheta =90^{\circ},45^{\circ},135^{\circ}$ and for kinetic energy release ranging from 5 to 50 eV ($\rm 0.39 \leq A_{CO} \leq 3.9$). }}
\end{figure}

For a qualitative understanding of the 3-body fragmentation, Fig.~\ref{fig:Simu_Dalitz} shows the Dalitz plots associated to the three intended geometrical configurations $\rm \uptheta =90^{\circ},45^{\circ},135^{\circ}$ and for KER ranging from 10 to 50 eV ($\rm 0.39 \leq A_{CO} \leq 3.9$). First, the figure indicates that, independently of the initial geometry, large kinetic energy released in the dissociation of the molecular $\rm CO^{2+}$ dication results in smaller energy of the $\rm CO^+$ ions. Indeed, the $\rm CO^+$ spends a shorter time in the field of the emitted $\rm C^+$ and $\rm O^+$ atomic ions and thus acquires less kinetic energy. Secondly, for a given KER of the $\rm CO^{2+}$ dication, the $\rm CO^+$ acquires a significantly larger kinetic energy in the case of the C-bonded or O-bonded ($\rm \uptheta =45^{\circ}$ or $\rm 135^{\circ}$) geometries because the $\rm CO^+$ ion is pushed away by the $\rm C^+$ or $\rm O^+$ ion emitted towards him. Finally, the dots associated to the $\rm \uptheta=45^{\circ}$ initial orientation are located on the right side of the points for $\rm \uptheta=90^{\circ}$ because the $\rm O^+$ atomic ion experienced a smaller Coulomb repulsion due to its central position within the fragmenting dimer. A similar argumentation justifies that the dots for $\rm \uptheta=135^{\circ}$ stand on the left side of the Dalitz plot. Comparing with the experimental data, where a large majority of events are located on the left part of the Dalitz plot (Fig.~\ref{fig:Dalitz}), the calculated $\rm 45^{\circ}$ (O-bonded) structure seems to be excluded since it does not match the experimental momentum repartition. On the other hand, both the $\rm 135^{\circ}$ (C-bonded) and $\rm 90^{\circ}$ structures may reproduce the experiment.

\subsubsection{Convolution with the experimental conditions}
To perform a more realistic comparison, the simulation algorithm has been extended in order to include most of the ingredients of the experimental conditions that may impact the data. The main causes responsible for the larger experimental distributions are the vibrational and rotational motion inside the dimer target and, to a smaller extent, the instrumental resolution of the spectrometer. To reproduce as closely as possible the experimental conditions, the 3-body dissociation is now simulated in the electric field of the recoil ion spectrometer and the ions are collected up to the detector. Events are generated using a Monte-Carlo approach. For each dissociation channel, the number of events is taken close to the statistics measured in the experimental data: respectively 2000, 1700, 500 and 500 events for the (1,[1,1]), (1,[1,2]), (1,[2,1]) and (1,[2,2]) channels. For each event, the initial conditions are set randomly and the ions are then propagated up to the detector. In such a way, the simulated data are created as a set of coincident detection of the time-of-flight, Y and Z positions on the detector and are further treated using the same analyzing procedure as for the experimental data, ensuring a fair comparison between experiment and calculation. 

Here is a detailed description of the parameters used in the model. First, the initial positions of the dimer were generated using random three dimensional Gaussian distributions with widths (FWHM) $\rm \Delta x=\Delta y=0.6 mm$ corresponding to the diameter of the ion beam collimator and $\rm \Delta z=1.5 mm$ according to the size of the gas jet. The initial velocities of the center of mass of the dimer is dominated by the momentum transferred to the target by the highly charged projectile ion during the collision [\onlinecite{IskandarPRA2018}]. These velocity distributions have been inferred from the experimental data and are found to be $\rm \Delta v_x=\Delta v_y=450 m/s$ (FWHM) in the transverse plane and about $\rm \Delta v_z = 45 m/s$ (FWHM) in the longitudinal direction. Then, the initial geometry of the dimer is fixed and the initial position of the point-like $\rm C^+$, $\rm O^+$ and $\rm CO^+$ ions are defined as described in the previous section. The orientation of the dimer with respect to the laboratory frame is then chosen according to an isotropic distribution. With this last assumption, we neglect any angular dependence of the electron capture process leading to these fragmentation channels. Such an angular dependence will however have no influence on the present determination of the geometrical structure of the dimer. To crudely account for the vibration of the molecules inside the dimer, a jitter of $\rm \Delta R_e=0.9~\text{\AA}$ (FWHM) on the intermolecular distance is used as deduced from the width of the KER spectra of the 2-body $\rm (CO)_2^{2+} \rightarrow CO^+ + CO^+$ channel. Additionally, an arbitrary jitter of $\rm \Delta \uptheta=20^{\circ}$ (FWHM) is used around the mean angles $\rm \uptheta=90^{\circ}$, $\rm 45^{\circ}$, $\rm 135^{\circ}$, $\rm 0^{\circ}$, $\rm 180^{\circ}$. Finally, to reproduce faithfully the experimental KER distribution, we randomly sample the experimental KER spectra associated to each channel and scale the effective repulsion coefficient $\rm A_{CO}$ between the two $\rm C^{p+}$ and $\rm O^{q+}$ accordingly. 

The coupled motion of each ion is then integrated up to the detector assuming a pure concerted fragmentation and accounting for the homogeneous extraction field E=40 V/m of the spectrometer. Finally, the detector response function has also been included using Gaussian distributions of $\rm \Delta TOF=1~ns$ for the timing resolution and of $\rm \Delta Y=\Delta Z=0.3~mm$ for the position resolution.

\section{COMPARISON TO EXPERIMENT}

\subsection{3-body channels}

\begin{figure*}[htpb]
\includegraphics[scale=0.51]{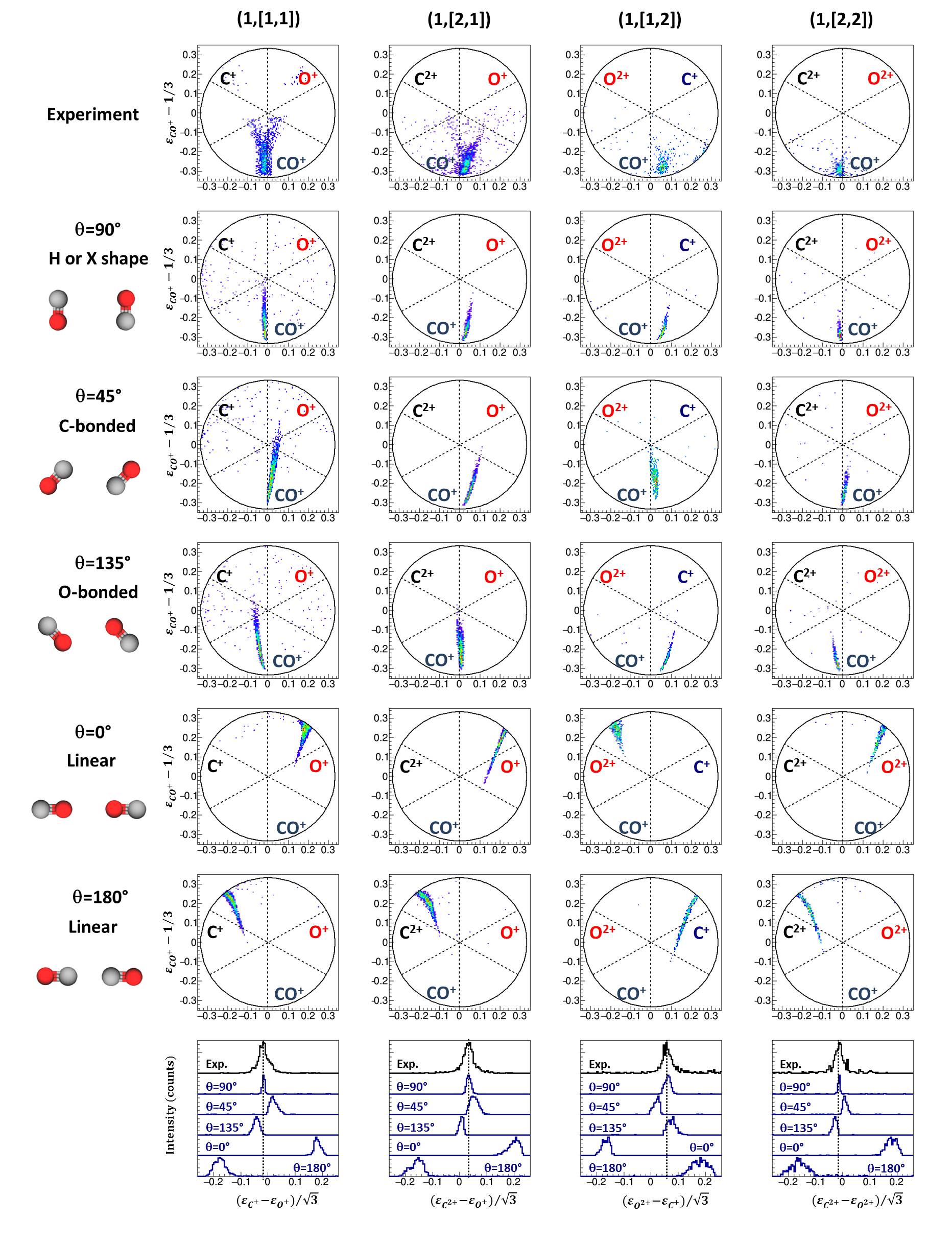}
\caption{\label{fig:Exp_simu_Dalitz}{Experimental and simulated Dalitz plots for the (1,[1,1]), (1,[2,1]), (1,[1,2]) and (1,[2,2]) channels. For the experimental (1,[1,1]) channel, only the concerted events are shown. For each channel, simulations have been run with initial orientations of $\rm \uptheta=90^{\circ},45^{\circ},135^{\circ}, 0^{\circ}, 180^{\circ}$ assuming a concerted fragmentation. Bottom : projection of the horizontal Dalitz coordinate $\rm x_D=\frac{\varepsilon_1 - \varepsilon_2}{\sqrt{3}}$. The vertical dashed line indicates the centroid of the experimental distribution for each fragmentation channel. }}
\end{figure*}

Simulations have been performed for the four fragmentation channels (1,[1,1]), (1,[2,1]), (1,[1,2]), (1,[2,2]). For each channel, simulations were performed for five initial angular structures of the dimer $\rm \uptheta=90^{\circ}$, $\rm 45^{\circ}$, $\rm 135^{\circ}$, $\rm 0^{\circ}$, $\rm 180^{\circ}$ corresponding to different conformations predicted by theoretical calculations. The resulting Dalitz plots are summarized on Fig.~\ref{fig:Exp_simu_Dalitz} where they are compared to the experimental data. For several channels and initial geometries of the simulation, few events appear as a diffuse background superimposed to the main contribution. These events are due to an incorrect reconstruction during the data analysis procedure and originates from the overlap of the time-of-flight distributions of the $\rm C^+$ and $\rm O^+$ or the $\rm C^+$ and $\rm O^{2+}$ partners due to their close mass-over-charge ratios. A small portion of these events are not efficiently filtered out by the momentum conservation law restriction imposed in the data analysis. They will be ignored in the forthcoming discussion. 

The bottom graphs on Fig.~\ref{fig:Exp_simu_Dalitz} represent the projection of the horizontal Dalitz coordinate $\rm x_D=\frac{\varepsilon_1 - \varepsilon_2}{\sqrt{3}}$ and gives the momentum repartition between the two atomic ions $\rm C^{p+}$ and $\rm O^{q+}$. Such one dimensional histograms are more appropriate to perform a quantitative comparison between experiment and calculations, especially when the statistic is low. Fig.~\ref{fig:Exp_simu_Dalitz} will now be used to deduce the three-dimensional structure of the dimer through a detailed comparison to simulated events. Even if the sensitivity to the initial angle $\rm \uptheta$ is limited, the combined information extracted from the four different fragmentation channels enables identifying the initial geometrical structure of the dimer. 
\begin{itemize}[label=$-$]
	\item the simulated Dalitz spectra for $\rm \uptheta =90^{\circ}$ and $\rm \uptheta =135^{\circ}$ are quite similar except for the (1,[2,1]) channel where significantly different distributions are observed. A close comparison of the $\rm x_D$ distributions clearly shows a better agreement of the experimental data with the $\rm \uptheta =90^{\circ}$ structure. For all fragmentation channels, the centroid of the simulated $\rm \uptheta =90^{\circ}$ data matches almost perfectly with the experiment while the $\rm \uptheta =135^{\circ}$ curves are significantly shifted to the right or left side of the measured value. 
	\item the simulated O-bonded structure ($\rm \uptheta =45^{\circ}$) does not reproduce the experimental data. It confirms our previous conclusion obtained from the 2-body dissociation (1,1) and (1,2) channels for which the measured dimer bond length does not match with theoretical predictions of the O-bonded structure.
	\item simulated data corresponding to $\rm \uptheta =0^{\circ}$ and $\rm 180^{\circ}$ show very specific Dalitz distributions for all fragmentation channels. The $\rm CO^+$ ion acquires a large amount of kinetic energy (events are located in the upper part of the Dalitz diagram) because of the Coulomb repulsion between the $\rm CO^+$ ion and the $\rm C^{p+}$ ($\rm \uptheta = 180^{\circ}$) or $\rm O^{q+}$ ($\rm \uptheta = 0^{\circ}$) ion emitted along the dimer axis. Such events have a very weak contribution in the experimental data, thus excluding a linear geometry of the dimer.
	\item T-shaped geometries are associated to two distinct $\rm \uptheta$ values ($\rm \uptheta \approx 0/180^{\circ}$ and $\rm \uptheta \approx 90^{\circ}$) [\onlinecite{HanJMS1997}]. They would result in two distinct contributions in the Dalitz plot depending on either the molecule perpendicular or aligned with the dimer axis breaks. The relative amplitude of these two contributions is expected to be about 50 \% because the electron capture probability on each molecular site of the dimer does not depend much on its relative orientation with respect to the dimer axis and that no rapid charge transfer between the two molecules is expected before fragmentation. As already discussed, very few events corresponding to $\rm \uptheta = 0$ or $\rm 180^{\circ}$ are observed in the experiment thus excluding a large contribution of T-shaped geometries.
\end{itemize}

The above detailed analysis of the 3-body fragmentation channels seems to reveal that the CO dimer structure consists mainly of two molecules oriented perpendicularly to the dimer axis ($\rm \uptheta =90^{\circ}$). The weak "V shape" observed in the (1,[1,1]) and (1,[2,1]) channels seems to arise from a small contribution of geometries in which the fragmenting molecule is aligned with the dimer axis. However, for $\rm \uptheta = 90^{\circ}$, the simulation does not reproduce the experimental widths of the Dalitz distributions which may indicate that the molecular orientation in the dimer $\rm \uptheta$ has a wider distribution than accounted for in the simulation. 

Surprisingly our data show no evidence for the C-bonded or O-bonded structures previously predicted by the calculations and associated to the two isomers observed in infrared spectroscopy measurements [\onlinecite{BrokesJChemPhys1999},\onlinecite{Surin2003},\onlinecite{Rezaei2013}]. As discussed in [\onlinecite{BrokesJChemPhys1999}], these two isomers are not fully distinct but rather overlapping and are associated to a wide angular extent. Transitions between these two conformations have also been observed in infrared spectroscopy presuming for a wide range of angular orientation ranging from C-bonded to O-bonded structure. Similar behaviors could also be deduced from the calculated PES of the dimer due to the low barrier between the two main minima. Vibrational excitation allows a dis-rotatory motion from the global C-bonded minimum towards the local O-bonded minimum while passing through parallel conformations. However according to the PES calculation, the rotatory motion is supposed to be accompanied by a stretching of the dimer because each angular orientation is associated to a different intermolecular distance. 

\begin{figure*}[htpb]
\includegraphics[scale=0.6]{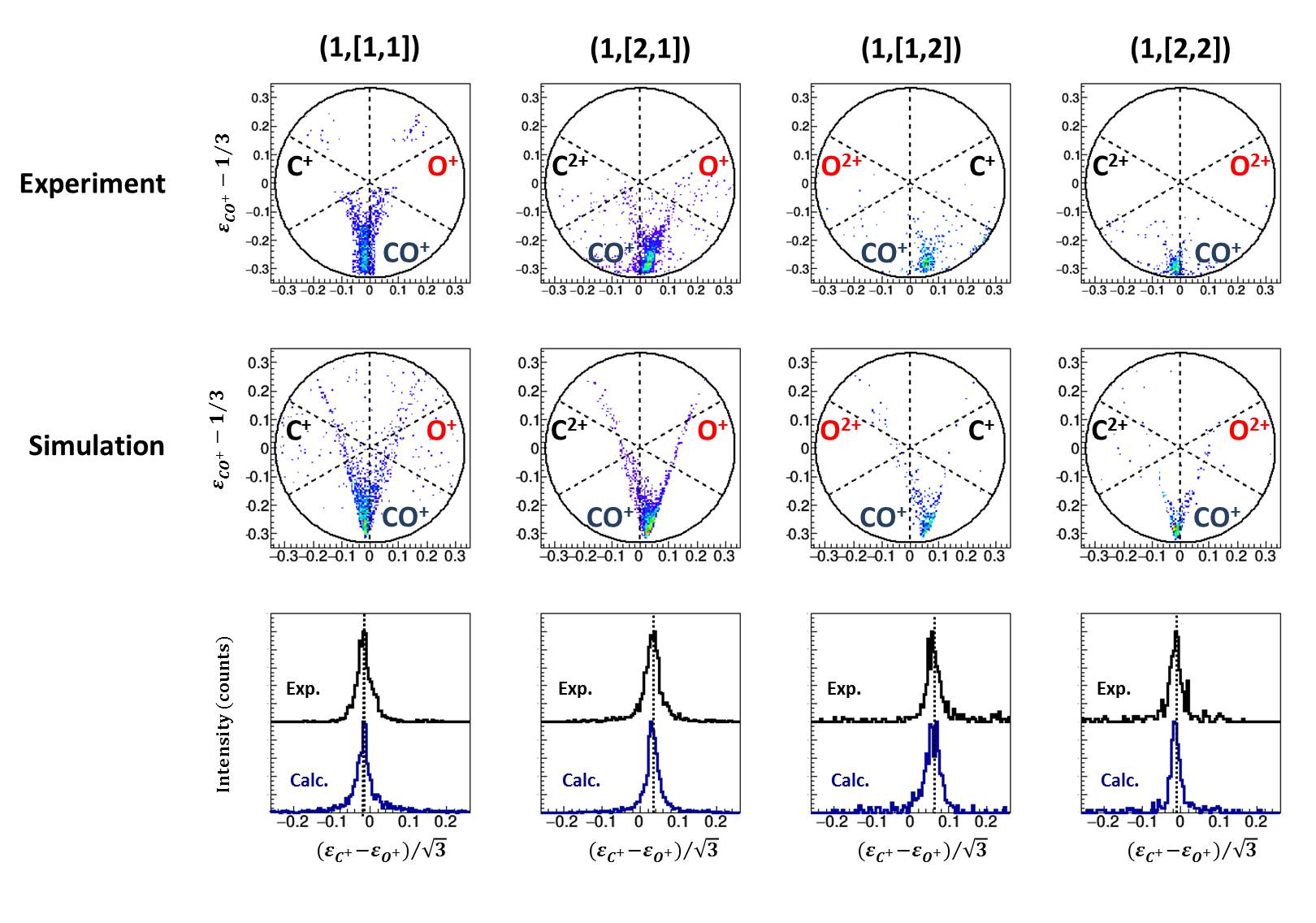}
\caption{\label{fig:Exp_simu_Dalitz_random}{Experimental and simulated Dalitz plots for the (1,[1,1]), (1,[2,1]), (1,[1,2]) and (1,[2,2]) channels. For the experimental (1,[1,1]) channel, only the concerted events are shown. For each channel, simulations have been run with a random initial orientation and assuming a concerted fragmentation. Bottom : projection of the horizontal Dalitz coordinate $\rm x_D=\frac{\varepsilon_1 - \varepsilon_2}{\sqrt{3}}$. The vertical dashed line indicates the centroid of the experimental distribution for each fragmentation channel. }}
\end{figure*}

As already mentioned, the actual vibrational temperature of the gas jet could be as large as 300~K (about $\rm 200~cm^{-1}\approx 25~meV$). The amplitude of the corresponding motion of each CO molecule in the dimer may be very wide [\onlinecite{BrokesJChemPhys1999}] and obviously much larger than the arbitrary jitter of $\rm \Delta \uptheta=20^{\circ}$ used in the previous simulations. A precise description of this motion and of the corresponding dimer conformations is by far out of the scope of our classical model. For the sake of simplicity, we ran a new set of simulations in the most extreme approximation of a fully random initial orientation of the two molecules with respect to the dimer axis. This approximation includes inherently a mixture of any of the previous initial angles ($\rm \uptheta=90^{\circ}$, $\rm 45^{\circ}$, $\rm 135^{\circ}$, $\rm 0^{\circ}$, $\rm 180^{\circ}$) but with a relative weight given by the sine dependance of the solid angle i.e. with a largely favored contribution of angles close to $\rm 90^{\circ}$. The resulting Dalitz plots are shown on Fig.~\ref{fig:Exp_simu_Dalitz_random} for the four fragmentation channels.

The experimental spectra are very well reproduced within such an approximation of a fully random initial orientation of the two CO molecules. The centroids of the Dalitz projections are similar to the ones obtained previously for $\rm \uptheta =90^{\circ}$ but the widths are significantly increased due to the contribution of other angles. It also confirms that the "V shape" is associated to initial angles close to $\rm \uptheta \cong 0$ or $180^{\circ}$. The very nice quantitative agreement of the simulated events with our data indicates that, rather than a well defined orientation, the molecules have a quasi-random orientation in the dimer. This observation is partly consistent with the previous results of infrared spectroscopy where the two isomers were also assumed to have a wide angular distribution. Moreover, our supersonic expansion method for the production of CO dimers is similar to the one used in [\onlinecite{BrokesJChemPhys1999}] and similar jet temperatures are thus expected. Even if the translational and rotational degrees of freedom are efficiently cooled down in the expansion, the vibrational temperature is large enough to allow highly excited vibrational states and results in a very large angular distribution. Within our experimental resolution, the vibrational motion is so large that it is compatible with a fully random orientation of the molecules inside the dimer.

\subsection{2-body channels}
As shown in section III.A, the dimer intermolecular distance can be deduced from KER spectra of the 2-body (1,1) and (2,1) channels. The pure Coulomb model gives about $R_e=\rm 4.5~\text{\AA}$ for the two channels but the value has to be corrected from the rotational energies of the two emitted molecular ions. During the dimer dissociation, a rotational motion of both ions is induced by the repulsion between the two dipoles of the $\rm CO^+$ and/or $\rm CO^{2+}$ ions. After dissociation, the resulting final rotational frequency depends on the initial conformation of the dimer (bond length and angular orientation). According to the previous analysis of the 3-body channels, the initial molecular orientation in the dimer is almost random and we used a Monte Carlo approach to average over all geometrical structures.

We thus adapted the previous model and extend it to account for the molecular rotational motion in the 2-body dissociation. The four atomic centers of the dimer are assumed to be point like particles. The partial charge on the carbon and oxygen atoms inside the $\rm CO^+$ ions are deduced from the electrostatic dipole moment of the $\rm CO^+$ ion. We used a repartition of respectively $\rm 90\%$ and $\rm 10\%$ on the carbon and oxygen atoms inside the $\rm CO^+$ molecular ion. For the (2,1) channel, we assumed the charge to be equally shared between the two atoms as we found no data on the electrostatic dipole moment of the metastable $\rm CO^{2+}$ dication in the literature. A pure Coulomb repulsion is used between each atomic ions from two different molecular sites (no intramolecular repulsion) and the binding intramolecular force is described by a restoring force proportional to the displacement along the molecular axis. The spring constant is chosen to reproduce vibrational wave numbers of about $\rm 2200~cm^{-1}$. This constant has very little influence on the results as only a negligible part of the initial energy is converted into intramolecular vibrational energy. Finally, the initial structure of the dimer is sampled using a Monte Carlo method by using a dimer bond length of $\rm 4.5~\text{\AA}$ with a jitter of $\rm 0.9~\text{\AA}$ (FWHM) and assuming a pure random orientation of the two molecules. 

This numerical integration of the ion motions is much more time consuming than in our previous model of the 3-body dissociation because of the fast vibrational motion of the two molecular ions. As a consequence, the ions are not propagated up to the detector but the calculation is stopped after $\rm 50 ps$ where we consider that the Coulomb repulsion between the two molecular ions vanishes.
After dissociation, the resulting rotational frequency ranges from 0 to about 2.5 $\rm ps^{-1}$ depending mainly on the initial orientation of the molecules with respect to the dimer axis. For each event, the total rotational energy of the two ions is calculated as the difference between the initial electrostatic potential energy of the system and the sum of the kinetic energy of the center-of-mass of the two molecular ions. For the (1,1) channel, we found that in average about 0.25 eV of the initial energy is converted into rotational energy. The experimental KER spectra is thus underestimated by 0.25 eV as our KER measurement includes only the energy of the center-of-mass of each $\rm CO^+$ molecular ion. Adding this rotational energy to the experimental KER leads to a total energy of 3.46 eV which corresponds to an equilibrium intermolecular distance of the CO dimer of $\rm R_e^{(1,1)}=4.17~\text{\AA}$. The uncertainty on the average rotational energy (0.25 eV) is hard to quantify but we estimate that its contribution to the error bars on $\rm R_e$ is about $\rm 0.2~\text{\AA}$ (FWHM). A similar method has been applied to the (2,1) channel but assuming an equal charge sharing between the two atoms in the $\rm CO^{2+}$ dication. We found that the rotational motion of the two molecular ions carries an energy of about 0.36 eV for this channel. The corrected total energy is then 6.73 eV and the corresponding bond length is about $\rm R_e^{(2,1)}= 4.3~\text{\AA}$ but with a large uncertainty due to the lack of information on the $\rm CO^{2+}$ dipole moment. As already discussed, the bond length derived from the two channels fits very well with the predicted true ground state of the CO dimer.

According to theoretical calculations, the minimum energy of each angular conformation is reached for different dimer bond length ranging from $\rm 3.6$ (O-bonded) to $\rm 4.3~\text{\AA}$ (C-bonded). We thus could expect very large KER distributions in the 2-body channels while the widths of our KER peaks (Fig.~\ref{fig:KER_2bodie}) do not show any significant contribution of the shortest bond length. Our interpretation of the 2-body and 3-body channels seems then not fully consistent with the shape of the PES of the CO dimer. This apparent contradiction may arise from the production of high excited vibrational states in the dimer gas jet. The high vibration excitation may partially destroy the correlation between angular orientation and intermolecular separation predicted in the calculations as we observe mainly the longest bond length corresponding to the ground state C-bonded conformation.

\section{CONCLUSION}
We measured the 2-body and 3-body dissociation of CO dimers following multiple electron capture by slow highly charged $\rm Ar^{9+}$ ions. We employed a Coulomb explosion imaging approach based on the axial-recoil approximation to retrieve the initial three dimensional structure of the dimer from the coincident measurement of the momentum vector of each emitted ions. Molecular orientation inside the dimer has been probed using the concerted fragmentation events of four different 3-body channels (1,[p,q]) and compared to 3-body simulated events assuming different initial geometries. This experimental method combined with a dedicated classical trajectory model enabled to get a direct and unambiguous determination of the dimer geometry. The structure of the CO dimer is found to be very diffuse and the measured Dalitz plots are found to be compatible with a purely random orientation of each CO molecule in the dimer. This result is somewhat consistent with infrared spectroscopy measurements as it confirms the very wide angular extent of the dimer conformation that may be attributed to a high vibrational temperature of the gas jet. According to the calculated PES, a dis-rotatory path covering both C-bonded and O-bonded geometries as well as intermediate parallel conformations could be responsible for such a large angular distribution at this temperature. The mean intermolecular distance has also been deduced from the KER distribution of the 2-body channels. After correction from the rotational energy acquired during dissociation, we found a bond length of about $\rm R_e=4.2~\text{\AA}$. This distance corresponds to the predicted true ground state of the CO dimer with a planar C-bonded geometry. Unexpectedly we found no evidence of the second isomer associated to a much shorter bond length and previously observed in infrared spectroscopy measurements. The relatively high temperature of the gas jet may be responsible for the presence of highly excited vibrational states for which the wave packet is centered around the global minimum of the PES (about $\rm 4.2~\text{\AA}$) but with a quasi random orientation of the two CO molecules.

\section{ACKNOWLEDGMENTS}
Experiment was performed at Grand Acc\'el\'erateur National d\textquotesingle Ions Lourds (GANIL) by means of CIRIL Interdisciplinary Platform, part of CIMAP laboratory, Caen, France. The authors want to thank the CIMAP and GANIL staff for their technical support.

\nocite{*}

\bibliography{CO-dimers_CEI_A-Mery}

\providecommand{\noopsort}[1]{}\providecommand{\singleletter}[1]{#1}%
\begin{thebibliography}{25}%
\makeatletter
\providecommand \@ifxundefined [1]{%
 \@ifx{#1\undefined}
}%
\providecommand \@ifnum [1]{%
 \ifnum #1\expandafter \@firstoftwo
 \else \expandafter \@secondoftwo
 \fi
}%
\providecommand \@ifx [1]{%
 \ifx #1\expandafter \@firstoftwo
 \else \expandafter \@secondoftwo
 \fi
}%
\providecommand \natexlab [1]{#1}%
\providecommand \enquote  [1]{``#1''}%
\providecommand \bibnamefont  [1]{#1}%
\providecommand \bibfnamefont [1]{#1}%
\providecommand \citenamefont [1]{#1}%
\providecommand \href@noop [0]{\@secondoftwo}%
\providecommand \href [0]{\begingroup \@sanitize@url \@href}%
\providecommand \@href[1]{\@@startlink{#1}\@@href}%
\providecommand \@@href[1]{\endgroup#1\@@endlink}%
\providecommand \@sanitize@url [0]{\catcode `\\12\catcode `\$12\catcode
  `\&12\catcode `\#12\catcode `\^12\catcode `\_12\catcode `\%12\relax}%
\providecommand \@@startlink[1]{}%
\providecommand \@@endlink[0]{}%
\providecommand \url  [0]{\begingroup\@sanitize@url \@url }%
\providecommand \@url [1]{\endgroup\@href {#1}{\urlprefix }}%
\providecommand \urlprefix  [0]{URL }%
\providecommand \Eprint [0]{\href }%
\providecommand \doibase [0]{http://dx.doi.org/}%
\providecommand \selectlanguage [0]{\@gobble}%
\providecommand \bibinfo  [0]{\@secondoftwo}%
\providecommand \bibfield  [0]{\@secondoftwo}%
\providecommand \translation [1]{[#1]}%
\providecommand \BibitemOpen [0]{}%
\providecommand \bibitemStop [0]{}%
\providecommand \bibitemNoStop [0]{.\EOS\space}%
\providecommand \EOS [0]{\spacefactor3000\relax}%
\providecommand \BibitemShut  [1]{\csname bibitem#1\endcsname}%
\let\auto@bib@innerbib\@empty
\bibitem [{\citenamefont {Brookes}(1999)}]{BrokesJChemPhys1999}%
  \BibitemOpen
  \bibfield  {author} {\bibinfo {author} {\bibfnamefont {M.~D.}\ \bibnamefont
  {Brookes}},\ }\href@noop {} {\bibfield  {journal} {\bibinfo  {journal} {J.
  Chem. Phys.}\ }\textbf {\bibinfo {volume} {111}},\ \bibinfo {pages} {7321}
  (\bibinfo {year} {1999})}\BibitemShut {NoStop}%
\bibitem [{\citenamefont {Surin}(2003)}]{Surin2003}%
  \BibitemOpen
  \bibfield  {author} {\bibinfo {author} {\bibfnamefont {L.}~\bibnamefont
  {Surin}},\ }\href@noop {} {\bibfield  {journal} {\bibinfo  {journal} {J. Mol.
  Spectrosc.}\ }\textbf {\bibinfo {volume} {222}},\ \bibinfo {pages} {93}
  (\bibinfo {year} {2003})}\BibitemShut {NoStop}%
\bibitem [{\citenamefont {Rezaei}(2013)}]{Rezaei2013}%
  \BibitemOpen
  \bibfield  {author} {\bibinfo {author} {\bibfnamefont {M.}~\bibnamefont
  {Rezaei}},\ }\href@noop {} {\bibfield  {journal} {\bibinfo  {journal} {J.
  Phys. Chem. A}\ }\textbf {\bibinfo {volume} {117}},\ \bibinfo {pages} {9612}
  (\bibinfo {year} {2013})}\BibitemShut {NoStop}%
\bibitem [{\citenamefont {Han}(1997)}]{HanJMS1997}%
  \BibitemOpen
  \bibfield  {author} {\bibinfo {author} {\bibfnamefont {H.}~\bibnamefont
  {Han}},\ }\href@noop {} {\bibfield  {journal} {\bibinfo  {journal} {J. Mol.
  Struct.}\ }\textbf {\bibinfo {volume} {418}},\ \bibinfo {pages} {1} (\bibinfo
  {year} {1997})}\BibitemShut {NoStop}%
\bibitem [{\citenamefont {Surin}(2007)}]{Surin2007}%
  \BibitemOpen
  \bibfield  {author} {\bibinfo {author} {\bibfnamefont {L.}~\bibnamefont
  {Surin}},\ }\href@noop {} {\bibfield  {journal} {\bibinfo  {journal} {J.
  Phys. Chem. A}\ }\textbf {\bibinfo {volume} {111}},\ \bibinfo {pages} {12238}
  (\bibinfo {year} {2007})}\BibitemShut {NoStop}%
\bibitem [{\citenamefont {Dawes}(2013)}]{Dawes2013}%
  \BibitemOpen
  \bibfield  {author} {\bibinfo {author} {\bibfnamefont {R.}~\bibnamefont
  {Dawes}},\ }\href@noop {} {\bibfield  {journal} {\bibinfo  {journal} {J.
  Phys. Chem. A}\ }\textbf {\bibinfo {volume} {117}},\ \bibinfo {pages} {7612}
  (\bibinfo {year} {2013})}\BibitemShut {NoStop}%
\bibitem [{\citenamefont {L\'egar\'e}(2005)}]{LegarePRA2005}%
  \BibitemOpen
  \bibfield  {author} {\bibinfo {author} {\bibfnamefont {F.}~\bibnamefont
  {L\'egar\'e}},\ }\href@noop {} {\bibfield  {journal} {\bibinfo  {journal}
  {Phys. Rev. A}\ }\textbf {\bibinfo {volume} {71}},\ \bibinfo {pages} {013415}
  (\bibinfo {year} {2005})}\BibitemShut {NoStop}%
\bibitem [{\citenamefont {Matsuda}(2007)}]{MatsudaJCP2007}%
  \BibitemOpen
  \bibfield  {author} {\bibinfo {author} {\bibfnamefont {A.}~\bibnamefont
  {Matsuda}},\ }\href@noop {} {\bibfield  {journal} {\bibinfo  {journal} {J.
  Chem. Phys.}\ }\textbf {\bibinfo {volume} {127}},\ \bibinfo {pages} {114318}
  (\bibinfo {year} {2007})}\BibitemShut {NoStop}%
\bibitem [{\citenamefont {Neumann}(2010)}]{NeumannPRL2010}%
  \BibitemOpen
  \bibfield  {author} {\bibinfo {author} {\bibfnamefont {N.}~\bibnamefont
  {Neumann}},\ }\href@noop {} {\bibfield  {journal} {\bibinfo  {journal} {Phys.
  Rev. Lett.}\ }\textbf {\bibinfo {volume} {104}},\ \bibinfo {pages} {103201}
  (\bibinfo {year} {2010})}\BibitemShut {NoStop}%
\bibitem [{\citenamefont {Wu}(2015)}]{WuJCP2015}%
  \BibitemOpen
  \bibfield  {author} {\bibinfo {author} {\bibfnamefont {C.}~\bibnamefont
  {Wu}},\ }\href@noop {} {\bibfield  {journal} {\bibinfo  {journal} {J. Chem.
  Phys.}\ }\textbf {\bibinfo {volume} {142}},\ \bibinfo {pages} {124303}
  (\bibinfo {year} {2015})}\BibitemShut {NoStop}%
\bibitem [{\citenamefont {Ulrich}(2011)}]{UlrichJPC2011}%
  \BibitemOpen
  \bibfield  {author} {\bibinfo {author} {\bibfnamefont {B.}~\bibnamefont
  {Ulrich}},\ }\href@noop {} {\bibfield  {journal} {\bibinfo  {journal} {J.
  Phys. Chem. A}\ }\textbf {\bibinfo {volume} {115}},\ \bibinfo {pages} {6936}
  (\bibinfo {year} {2011})}\BibitemShut {NoStop}%
\bibitem [{\citenamefont {Wu}(2014)}]{WuJCP2014}%
  \BibitemOpen
  \bibfield  {author} {\bibinfo {author} {\bibfnamefont {C.}~\bibnamefont
  {Wu}},\ }\href@noop {} {\bibfield  {journal} {\bibinfo  {journal} {J. Chem.
  Phys.}\ }\textbf {\bibinfo {volume} {140}},\ \bibinfo {pages} {141101}
  (\bibinfo {year} {2014})}\BibitemShut {NoStop}%
\bibitem [{\citenamefont {Gong}(2013)}]{GongPRA2013}%
  \BibitemOpen
  \bibfield  {author} {\bibinfo {author} {\bibfnamefont {C.}~\bibnamefont
  {Gong}},\ }\href@noop {} {\bibfield  {journal} {\bibinfo  {journal} {Phys.
  Rev. A}\ }\textbf {\bibinfo {volume} {88}},\ \bibinfo {pages} {013422}
  (\bibinfo {year} {2013})}\BibitemShut {NoStop}%
\bibitem [{\citenamefont {Wu}(2013)}]{WuPRL2013}%
  \BibitemOpen
  \bibfield  {author} {\bibinfo {author} {\bibfnamefont {C.}~\bibnamefont
  {Wu}},\ }\href@noop {} {\bibfield  {journal} {\bibinfo  {journal} {Phys. Rev.
  Lett.}\ }\textbf {\bibinfo {volume} {110}},\ \bibinfo {pages} {103601}
  (\bibinfo {year} {2013})}\BibitemShut {NoStop}%
\bibitem [{\citenamefont {McClelland}(1979)}]{McClellandJPC1979}%
  \BibitemOpen
  \bibfield  {author} {\bibinfo {author} {\bibfnamefont {G.}~\bibnamefont
  {McClelland}},\ }\href@noop {} {\bibfield  {journal} {\bibinfo  {journal} {J.
  Phys. Chem.}\ }\textbf {\bibinfo {volume} {8}},\ \bibinfo {pages} {947}
  (\bibinfo {year} {1979})}\BibitemShut {NoStop}%
\bibitem [{\citenamefont {Wall}(2016)}]{WallJPB2016}%
  \BibitemOpen
  \bibfield  {author} {\bibinfo {author} {\bibfnamefont {T.}~\bibnamefont
  {Wall}},\ }\href@noop {} {\bibfield  {journal} {\bibinfo  {journal} {J. Phys.
  B}\ }\textbf {\bibinfo {volume} {49}},\ \bibinfo {pages} {243001} (\bibinfo
  {year} {2016})}\BibitemShut {NoStop}%
\bibitem [{\citenamefont {Lundqvist}(1995)}]{Lundqvist1995}%
  \BibitemOpen
  \bibfield  {author} {\bibinfo {author} {\bibfnamefont {M.}~\bibnamefont
  {Lundqvist}},\ }\href@noop {} {\bibfield  {journal} {\bibinfo  {journal}
  {Phys. Rev. Lett.}\ }\textbf {\bibinfo {volume} {75}},\ \bibinfo {pages}
  {1058} (\bibinfo {year} {1995})}\BibitemShut {NoStop}%
\bibitem [{\citenamefont {Ding}(2017)}]{DingPRL2017}%
  \BibitemOpen
  \bibfield  {author} {\bibinfo {author} {\bibfnamefont {X.}~\bibnamefont
  {Ding}},\ }\href@noop {} {\bibfield  {journal} {\bibinfo  {journal} {Phys.
  Rev. Lett.}\ }\textbf {\bibinfo {volume} {118}},\ \bibinfo {pages} {153001}
  (\bibinfo {year} {2017})}\BibitemShut {NoStop}%
\bibitem [{\citenamefont {Kumar}(2010)}]{KumarJP2010}%
  \BibitemOpen
  \bibfield  {author} {\bibinfo {author} {\bibfnamefont {P.}~\bibnamefont
  {Kumar}},\ }\href@noop {} {\bibfield  {journal} {\bibinfo  {journal} {Pramana
  - J. Phys.}\ }\textbf {\bibinfo {volume} {74}},\ \bibinfo {pages} {49}
  (\bibinfo {year} {2010})}\BibitemShut {NoStop}%
\bibitem [{\citenamefont {Freidlander}(1932)}]{Friedlander1932}%
  \BibitemOpen
  \bibfield  {author} {\bibinfo {author} {\bibfnamefont {E.}~\bibnamefont
  {Freidlander}},\ }\href@noop {} {\bibfield  {journal} {\bibinfo  {journal} {Z
  Physik}\ }\textbf {\bibinfo {volume} {76}} (\bibinfo {year}
  {1932})}\BibitemShut {NoStop}%
\bibitem [{\citenamefont {Andersen}(1993)}]{Andersen1993}%
  \BibitemOpen
  \bibfield  {author} {\bibinfo {author} {\bibfnamefont {L.}~\bibnamefont
  {Andersen}},\ }\href@noop {} {\bibfield  {journal} {\bibinfo  {journal}
  {Phys. Rev. Lett.}\ }\textbf {\bibinfo {volume} {71}},\ \bibinfo {pages}
  {1812} (\bibinfo {year} {1993})}\BibitemShut {NoStop}%
\bibitem [{\citenamefont {Mrugala}(2008)}]{MrugalaJCP2008}%
  \BibitemOpen
  \bibfield  {author} {\bibinfo {author} {\bibfnamefont {F.}~\bibnamefont
  {Mrugala}},\ }\href@noop {} {\bibfield  {journal} {\bibinfo  {journal} {J.
  Chem. Phys.}\ }\textbf {\bibinfo {volume} {129}},\ \bibinfo {pages} {064314}
  (\bibinfo {year} {2008})}\BibitemShut {NoStop}%
\bibitem [{\citenamefont {Song}(2019)}]{SongPRA2019}%
  \BibitemOpen
  \bibfield  {author} {\bibinfo {author} {\bibfnamefont {P.}~\bibnamefont
  {Song}},\ }\href@noop {} {\bibfield  {journal} {\bibinfo  {journal} {Phys.
  Rev.A}\ }\textbf {\bibinfo {volume} {99}},\ \bibinfo {pages} {053427}
  (\bibinfo {year} {2019})}\BibitemShut {NoStop}%
\bibitem [{\citenamefont {Ben-Itzhak}(1993)}]{BenItzhakPRA1993}%
  \BibitemOpen
  \bibfield  {author} {\bibinfo {author} {\bibfnamefont {I.}~\bibnamefont
  {Ben-Itzhak}},\ }\href@noop {} {\bibfield  {journal} {\bibinfo  {journal}
  {Phys. Rev. A}\ }\textbf {\bibinfo {volume} {47}},\ \bibinfo {pages} {2827}
  (\bibinfo {year} {1993})}\BibitemShut {NoStop}%
\bibitem [{\citenamefont {Iskandar}(2018)}]{IskandarPRA2018}%
  \BibitemOpen
  \bibfield  {author} {\bibinfo {author} {\bibfnamefont {W.}~\bibnamefont
  {Iskandar}},\ }\href@noop {} {\bibfield  {journal} {\bibinfo  {journal}
  {Phys. Rev. A}\ }\textbf {\bibinfo {volume} {98}},\ \bibinfo {pages} {012701}
  (\bibinfo {year} {2018})}\BibitemShut {NoStop}%
\end{thebibliography}%

\end{document}